\newcommand{\gaas}{$\rm GaAs$} 
\newcommand{\algaas}{$\rm Al_{x}\-Ga_{1-x}\-As$}

\newcommand{\gaasalgaas}{$\rm GaAs/Al_{x}\-Ga_{1-x}\-As$}
\newcommand{\gaasingaas}{$\rm GaAs/In_{x}\-Ga_{1-x}\-As$}
\newcommand{\ingap}{$\rm InGaP_{x}$} 
\newcommand{\gaasp}{$\rm Ga_{x}\-AsP$}
\newcommand{\ingaasp}{$\rm InGa_{0.53x}\-As_{x}\-P$}

\newcommand{\voc}{$\rm V_{oc}$}

\newcommand{\figcap}[1]{\caption{\protect \footnotesize #1}}
\newcommand{\smalleps}[3]{\begin{figure}[#1]\centerline{ 
   \epsfxsize=8cm 
   \epsffile{#2}} 
   \figcap{#3} 
   \end{figure}} 

\documentclass[twocolumn]{alaska}
\usepackage{epsf,emlines,fleqn}

\begin{document}
\setlength{\paperwidth}{21cm} \setlength{\paperheight}{29.7cm}
\setlength{\topmargin}{0.4cm} \setlength{\oddsidemargin}{-0.4cm}

\pagestyle{empty}
\baselineskip 1em

\begin{frontmatter}

\title{\large \bf SIMULATING MULTIPLE QUANTUM WELL SOLAR CELLS }
\author[London]{James P. Connolly}
\author[London]{Jenny Nelson}
\author[London]{Keith W.J. Barnham }
\author[London]{Ian Ballard}
\author[London]{C.Roberts}
\author[Sheffield]{J.S. Roberts}
\author[nottingham]{C.T.Foxon}

\vspace{-0.1cm}
\address[London]{Blackett Laboratory, Imperial College of
Science, Technology and Medicine, London SW7 2BZ }
\address[CEMD]{Centre for Electronic Materials and Devices, Imperial
College, London SW7 2BZ UK}
\address[Sheffield]{ EPSRC III-V Facility, University of Sheffield, Sheffield S1 3JD UK}
\address[nottingham]{ University of Nottingham, Nottingham N67 2RD UK}

\thanks{Electronic mail: j.connolly@ic.ac.uk}

\vspace{-0.16cm}

\begin{keyword}
\vspace{-2\baselineskip}
    Quantum well solar cell, dark current, light current, efficiency
\end{keyword}

\end{frontmatter}

\section{abstract}

 The quantum well solar cell (QWSC) has been
proposed as a route to higher efficiency than that
attainable by homojunction devices.  Previous studies have established
that carriers escape the quantum wells with high efficiency in forward bias
and contribute to the photocurrent. 
Progress in resolving the efficiency limits of these cells
has been dogged by the lack of a theoretical model
reproducing both the enhanced carrier generation and enhanced
recombination due to the quantum wells.  Here we present a model which
calculates the incremental generation and recombination due to the QWs
and is verified by modelling the experimental light and dark
current-voltage characteristics of a range of III-V quantum well
structures. We find that predicted dark currents are significantly greater
than experiment if we use lifetimes derived from homostructure devices.
Successful simulation of light and dark currents can be obtained only by
introducing a parameter which represents a reduction in the quasi-Fermi level
separation.

\section{Introduction}\label{introduction}

The quantum well solar cell (QWSC) \cite{barnham90} consists of a
$p-i-n$ structure with nm-wide regions of low bandgap (quantum wells)
in the nominally undoped intrinsic region where carriers occupy
discrete energy levels.

Experiment has shown (\cite{paxman93}, \cite{barnes96}) that
photo-excited electrons and holes in these quantum wells escape with
high efficiency in forward bias and contribute to the photocurrent. 
If the photocurrent enhancement exceeds the dark current increase
this may be a highly efficient solar cell design.

Detailed balance arguments in the radiative limit \cite{Araujo94} have
suggested the \voc\ is determined by the lowest absorption energy which
is the well bandgap.  Further work \cite{voltageenhancement} however
indicates that efficiency enhancements are possible over single bandgap
designs in non-radiative dominated structures. This has been confirmed
by work (\cite{nelson96} \cite{Tsui96}) which showed smaller quasi-Fermi
level (QFL) separation in the wells than in the barriers. This suggests that
significant efficiency enhancements are possible with this solar cell design.

This paper describes QWSC photocurrent and dark current calculations with a
view to 
establishing whether the effect of enhanced photocurrent overcompensates
the dark current losses and secondly modelling the overall device performance.
The photocurrent is calculated from the diffusion
equation applied to photogenerated minority carriers in bulk and well
material and is well understood.  The dark current aspect is a
Shockley-Read-Hall (SRH) calculation including an ideal diode component. 
The model examines the extent to which the dark current behaviour of
QWSCs can be characterised in terms of the quantum well density of
states without considering escape and capture rates from the quantum
wells.  Furthermore, it explores the extent to which the QWSC
dark current is determined by the SRH recombination rates at the
centre of the intrinsic region, and for many wells where this
methodology is expected to be accurate.  This is explored by modelling
a range of experimetally measured dark currents in the well
characterised \gaasalgaas\ system.

\section{QWSC current voltage model \label{theory}}

The program "SOL" calculates the incremental light and dark currents
due to the quantum wells. It will be shown
that the changes in photocurrent and dark current can both be
reduced to functions of the quantum well densities of states.

The photocurrent model of the QWSC consists of a spectral response
calculation as a function of bias, cell dimensions and quantum well
energy spectrum. Cell photocurrents can then be
calculated for different incident light spectra and power levels. 
This has been described for the \gaasalgaas\, \gaasingaas\, \ingap\,
\gaasp\ and \ingaasp\ materials systems \cite{paxman93}, \cite{barnes96}.

The dark currents measured in QWSC structures in these materials show
idealities close to 2.  This is characteristic of 
SRH recombination in the depletion layer.  The dark current model
applied here is a simplified implementation of a self consistent model
by Nelson {\it et al} \cite{nelson99} which has succesfully described the
dark currents of single quantum well cells for a range of single well
positions and $i$ region background doping levels.

The carrier generation rate $G$ at wavelength $\lambda$, position $x$,
incident photon flux $F$, $R$ the surface reflectivity and
absorption coefficient $\alpha$ is defined as

\begin{equation}\label{generation}
     G(x,\lambda)=F(\lambda) (1-R(\lambda)) \alpha(x) 
e^{-\int_{0}^{x}\alpha x dx}
\end{equation}

In the quantum wells the generation is
calculated from Fermi's golden rule governing transition probabilities
between electron and hole well states.  These are found by solving the
effective mass equation for electrons and holes confined in the QW within
the envelope function approximation. Omitting constants for brevity it is
written as

\begin{equation}
\label{qwabsorption}
\alpha_{qw} \sim \sum_{jn=1,jp=1}^{N states}M^{2} H( {E_{jn}-E_{jp}-\hbar 
\omega})
\end{equation}

where the double sum is over initial and final valence and conduction
band states labelled $jn$ and $jp$, $M$ is the transition Hamiltonian,
$H$ is the Heaviside step function delta, $E_{jn}$ and $E_{jp}$ are
the energy levels of the initial and final quantum well states.

The photocurrent $J_{i}$ due to the QW at any $\lambda$ is found by
integrating $G$ across the QW layers and assuming 100\% escape. The
QW spectral response is added on to that of the bulk regions of the device
calculated as in \cite{paxman93}.

Figure 1 shows the calculated and measured QE for an
\algaas\ QWSC comprising 30 \gaas\ wells together with the same for
a control sample which is identical but without wells. Under an standard
AM1.5 spectrum
the theoretical short circuit current ($J_{sc}$) for
this sample is 108$Am^{-2}$ versus 116$Am^{-2}$
calculated from the experimental SR. These results agree with
calibrated results from NREL to within $3\%$ in the worst case.

\smalleps{htbp}{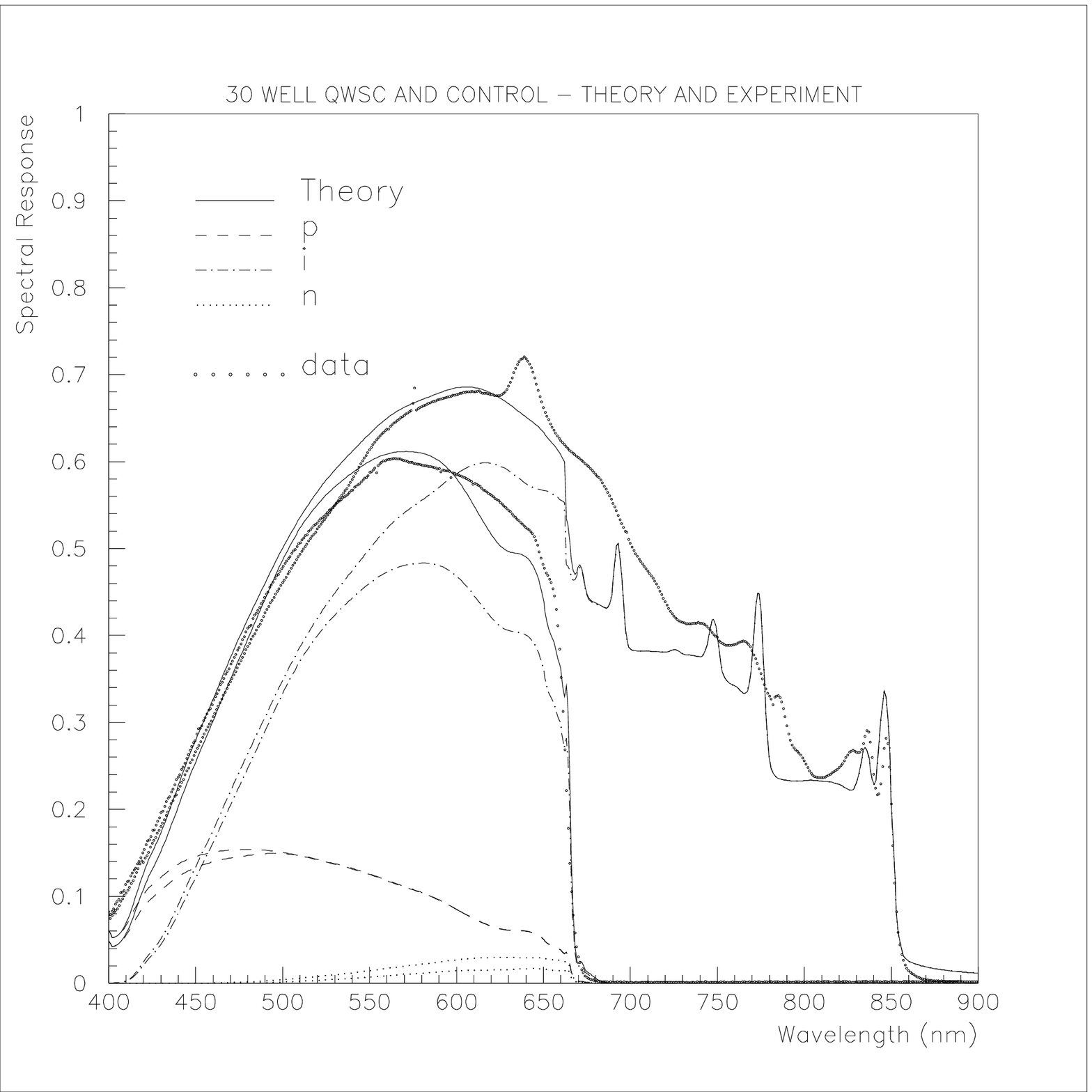}{Experimental and calculated SR of a 30well \algaas\
sample with aluminium $x=30\%$ \label{srexample}}

The SRH \cite{srh52} recombination rate $U(x)$ in the i region is
given by
\begin{equation}
U(x)=\frac
{pn-n_{i}^2}
{\tau_{n}(p+p_{t})+\tau_p(n+n_{t})}
\end{equation}

where where $p_{t}$ and $n_{t}$ are hole and electron populations in
trap levels in the bandgap, $p$ and $n$ are hole and electron
populations in the valence and conduction bands and $\tau_{p}$ and
$\tau_{n}$ are their respective lifetimes.

Electron and hole populations in barrier regions of the intrinsic
region are expressed in terms of the intrinsic population $n_{i}$,
Fermi levels $E_{f}$ and the intrinsic energy $E_{i}$

\begin{equation}
\begin{array}{l}
n_{b}=n_{i} e^{ \left( \frac{E_{fn}-E_{i}} {KT} \right)} \\
p_{b}=n_{i} e^{ \left( \frac{E_{i}-E_{fp}} {KT} \right)} \\
\end{array}
\end{equation}

where $n_{i}$ is given by the effective masses of electrons and holes
via the conduction and valence band effective densities of states and
the law of mass action in equilibrium, and  for $n_{b} $ the subscript
$b$ refers to barrier material.

It has been shown \cite{nelson99} that the variation in electron and
hole populations in well and barrier regions can be reduced to a
difference in effective mass within the parabolic band approximation
and a constant factor modifying the conduction and valence band
densities of states which is determined by a sum over quantum well
allowed states.  These are written in terms of two functions
$\theta_{p}$ and $\theta_{n}$.  Electron and hole populations $n_{q}$
and $p_{q}$ in the wells are then given as follows in terms of the barrier
concentrations

\begin{equation}
\begin{array}{l}
n_{q}=n_{b}e^{\theta_{n}} \\
p_{q}=p_{b}e^{\theta_{p}} \\
\end{array}
\end{equation}

The $\theta$ functions are as follows
\begin{equation}
\label{thetas}
\begin{array}{l}
\theta_{n}=k T \ 
\ln \left[ \frac{1}{L} (\frac{2} {N_{C}})^{1/3} \frac{m_{nq}}{m_{n}}
\sum_{jn=1}^{N}
e^{(\frac{E_{cq}-E_{jn}}{KT})} 
\right] \\

\theta_{p}=k T \ 
\ln \left[ \frac{1}{L} (\frac{2} {N_{V}})^{1/3} \frac{m_{pq}}{m_{p}}
\sum_{jp=1}^{N}
e^{(\frac{E_{jp}-E_{jq}}{KT})}
\right] \\
\end{array}
\end{equation}

where $L$ is the quantum well width. $m_{n,p}$ are electron and hole masses
in the barrier and  $m_{nq,pq}$ are the masses in the well.
These functions contain
all information necessary about the quantum wells with the sole
exception of carrier lifetimes.  We note that they are symmetrical to
the well photocurrent enhancement term of equation \ref{qwabsorption}.  Both
involve a sum over well states and the necessary information is the
same as in equation \ref{qwabsorption} and reduces to band structure
data about the well relative to the barriers.  The functions will not
be symmetrical for electrons and holes because of different
valence/conduction offsets and effective masses.

The contributions from neutral layers are  included in terms of the ideal
diode formalism  The electron and hole Fermi-levels
are set equal to the valence and conduction band edges.  The depletion
approximation is made with the result that photogenerated carriers are
omitted in the dark current calculation.

The dark current of the device in this picture is given by the integral
of the SRH recombination rate over the depletion region including p and n
layer depletion layers if present, incrememented by the ideal diode contribution
from the charge neutral regions. An example SRH profile is shown in 
figure \ref{shrexample}. The quantum wells show are step-like peaks in 
the profile, and clearly dominate the dark current of a QWSC.

\smalleps{htbp}{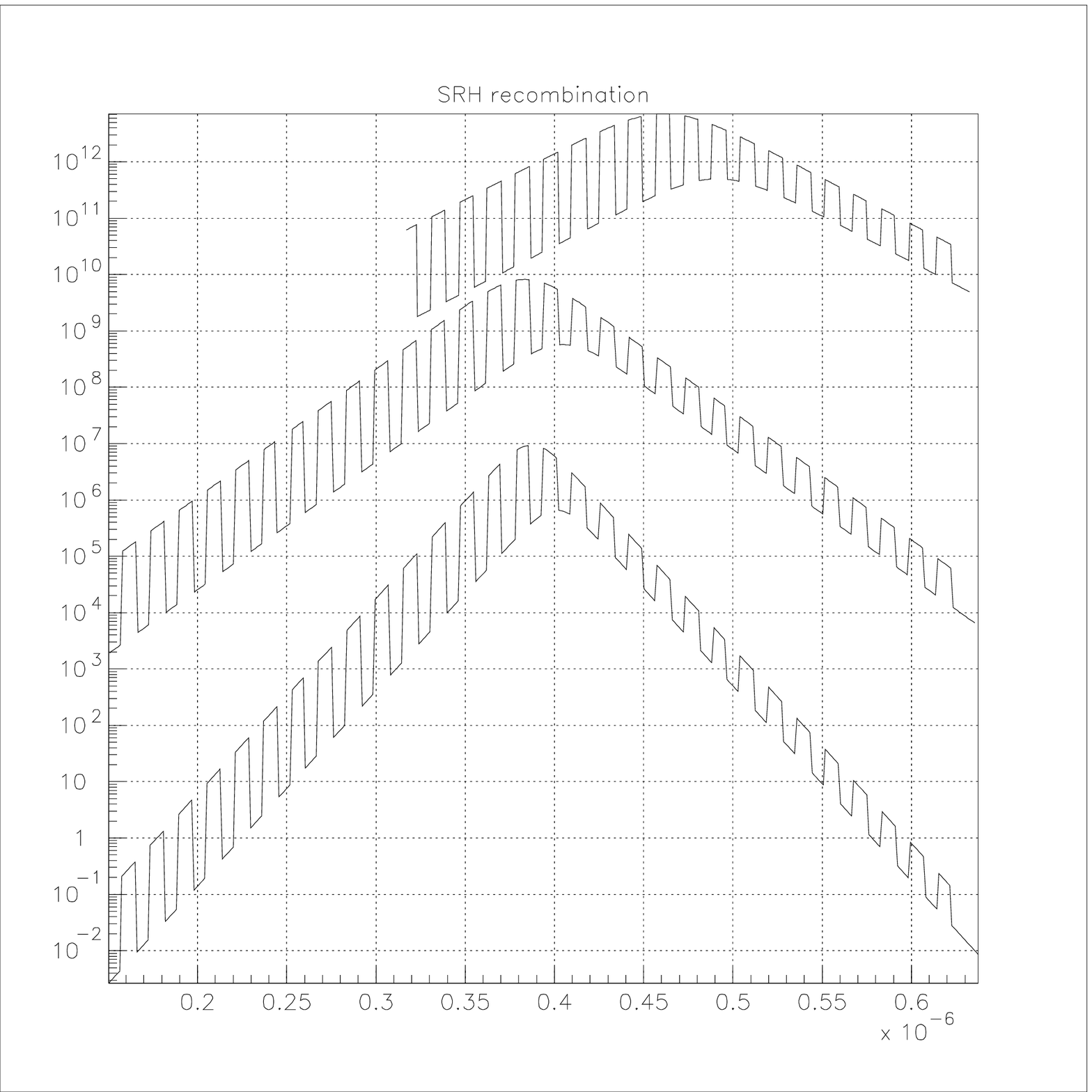}{SRH recombination profile for the 30
well QWSC of figure \ref{srexample} for a biasses 0.8V, 1.15 and 1.5V,
showing different $\theta_{n}$ and $\theta_{p}$ and the reduction of
SCR width at high bias.\label{shrexample}}

\section{Experiment and Modelling}

Samples were grown by MBE and MOVPE in the Centre for Electronic
Materials and Devices in Imperial College, Phillips Laboratories (Redhill)
and the Sheffield III-V Semiconductor Facility. They take the form of
$p-i-n$ diodes. Structures having \gaas\ wells in the $i$ region will be
referred to as QWSCs and identical $p-i-n$ structures of homogeneous
composition grown in the same growth run are termed controls. The
experimental configuration has been described previously \cite{paxman93}.

The fitting parameters for dark currents are the non radiative lifetimes.
Estimates for these are drawn from studies of barrier and well controls
which are \gaasalgaas\ and \gaas\ controls. These lifetimes are then applied
to QWSC structures and the resulting prediction compared with experiment.

Figure 3 shows modelled and calculated IV for \gaas\ and
\algaas\ controls.  The upper curves shows a \gaas\ control.  The
model shows SRH and ideal dark currents and their sum.  The onset of
radiatively dominated behaviour is visible below the built in
voltage of $1.44V$ and is well described. 
The parameters for the ideal Shockley fit are the same as those used for
the SR leaving the non radiative lifetimes as the only parameters.
They are chosen equal in this case.  Making them different shifts 
the SRH peak with position and changes the ideality of the dark
current in disagreement with experiment. The lifetime of 10ns is in close
agreement with lifetimes used in previous work \cite{nelson99}.

The lower curves shows similar data and theory for an \algaas\ control
with a much shorter SRH lifetime of 0.6ns. Again this is in agreement
with previous work by the same factor in a different sample.
Modelling an identical control with aluminium mole fractions $x=20\%$
yields a lifetime of 0.05ns.

\smalleps{htbp}{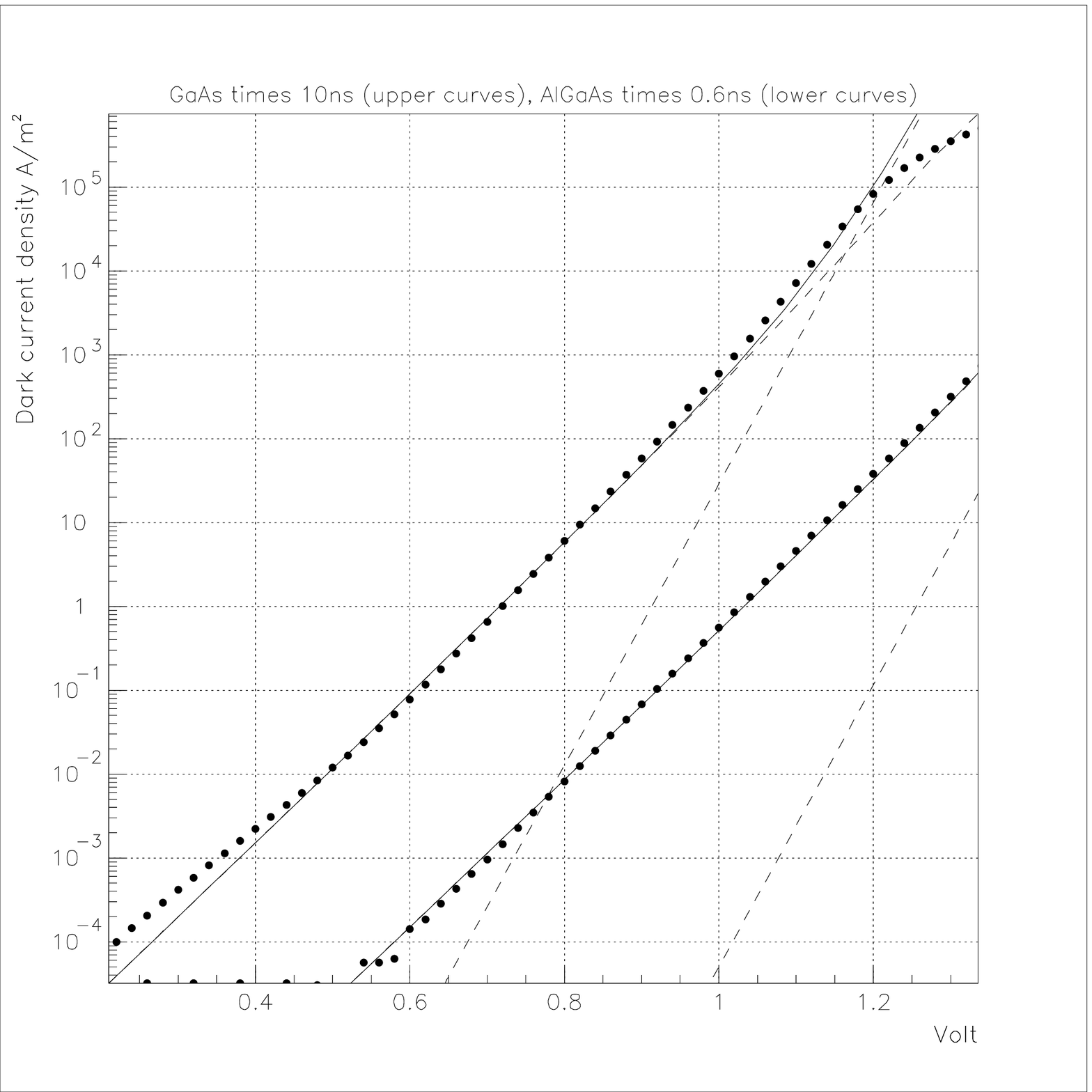}
{\gaas\ control dark current and lifetimes\label{gaasiv}}

\smalleps{htbp}{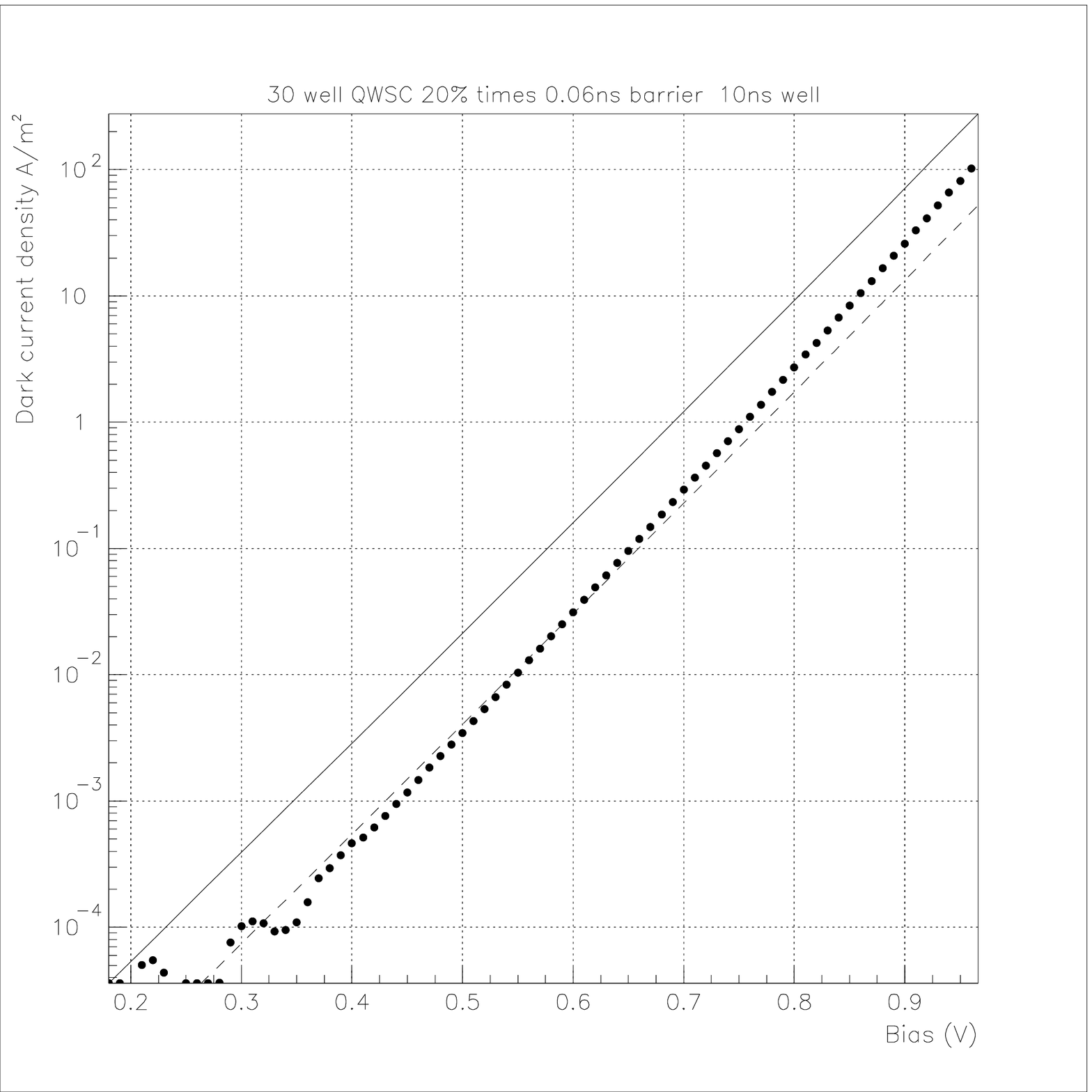}
{30 well QWSC Al 20\% (dashed line modified
$\delta E_{f}$ see text)\label{iv2027}}

\smalleps{htbp}{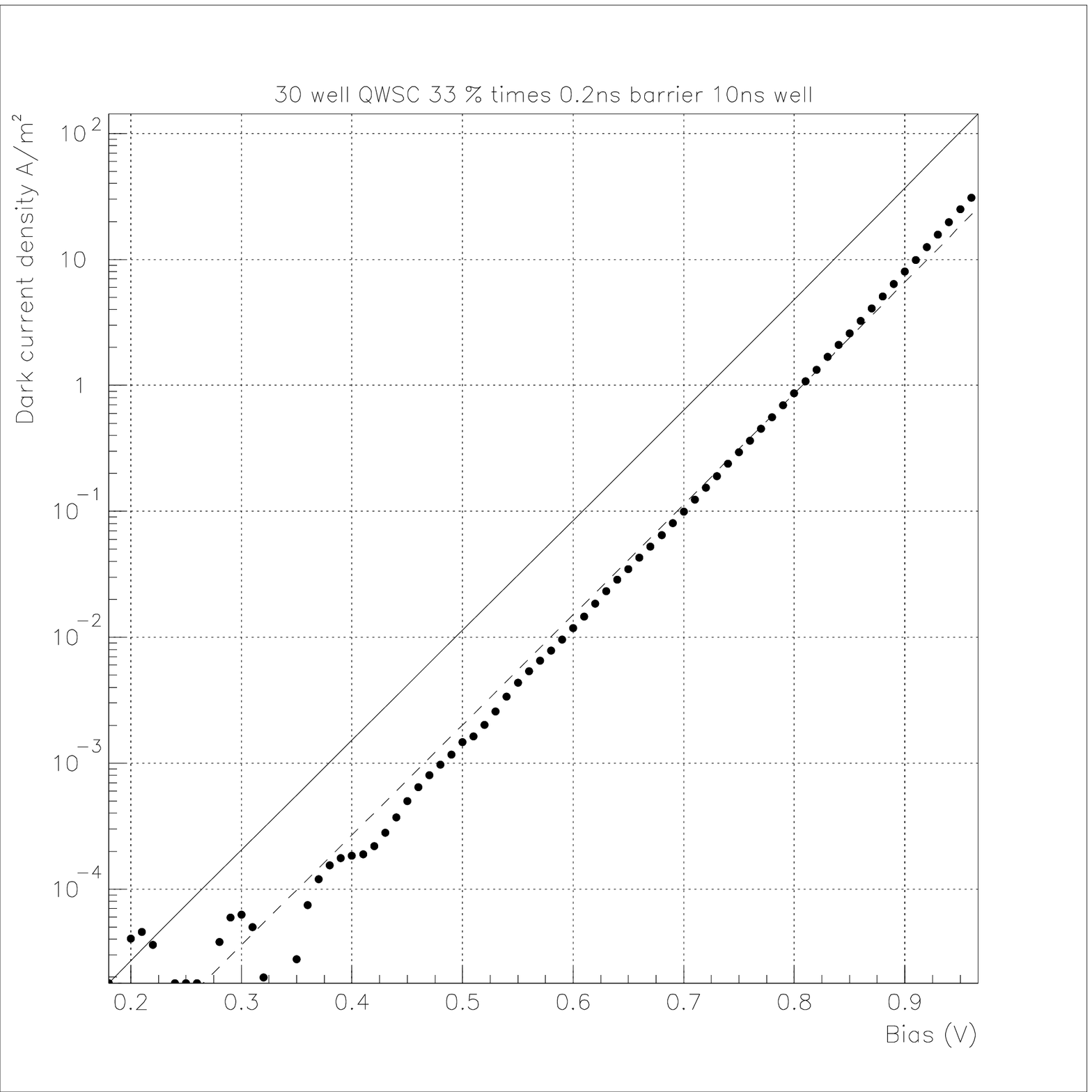}
{30 well QWSC Al 33\% (dashed line modified
$\delta E_{f}$ see text)\label{iv2029}}

\smalleps{htbp}{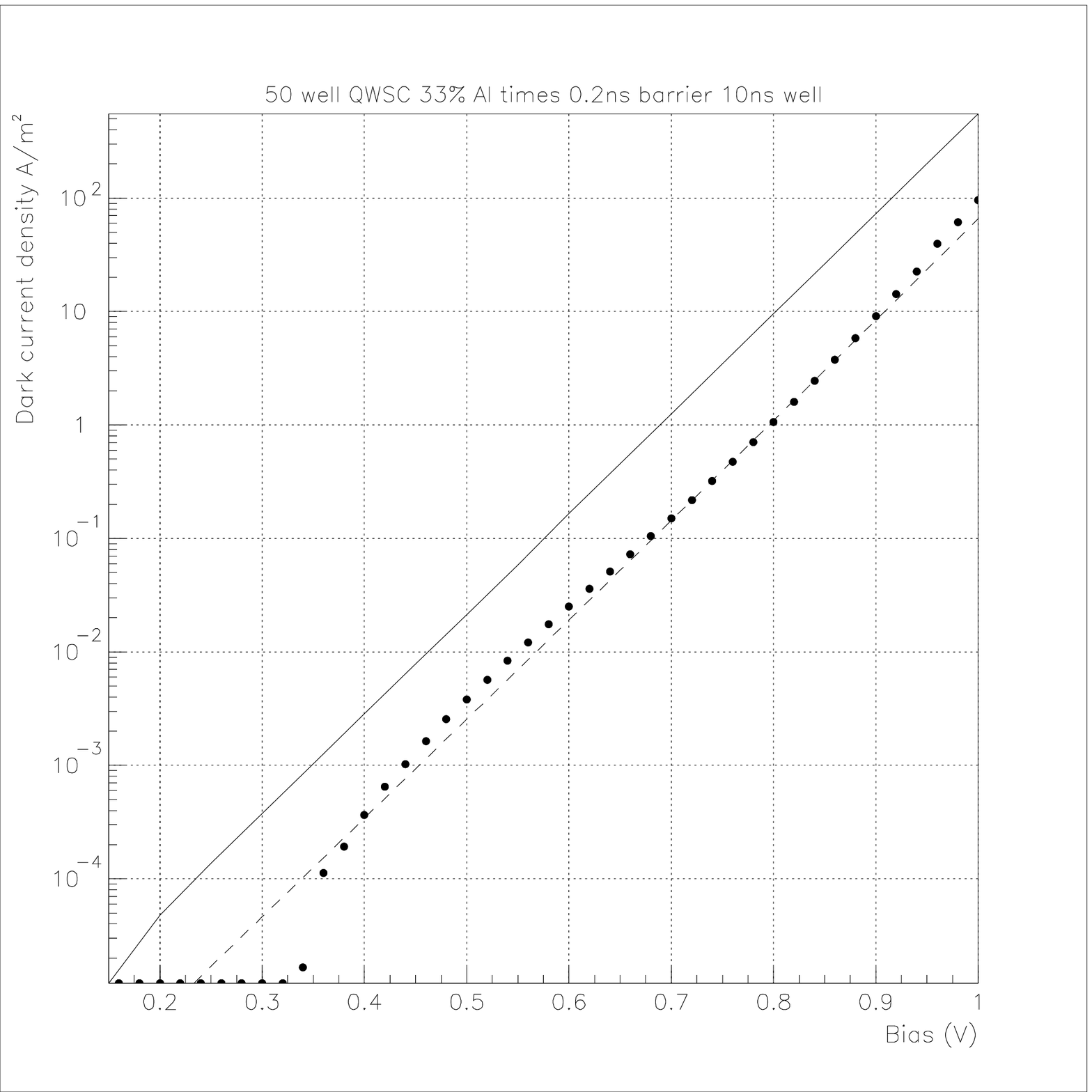}
{50 well QWSC Al 33\% (dashed line modified
$\delta E_{f}$ see text)\label{iv951}}


Figures \ref{iv2027} and \ref{iv2029}  show dark current modelling for
30 well MBE grown QWSCs with Al composition of 20\% and 33\% and with
lifetimes derived from relevant controls assuming similar material quality.

Figure \ref{iv951} shows the same data for a 50 well QWSC with Al
fractions 33\%. Fitting the data can again be achieved by reducing the
QFL separation by the same shift.

We have observed that experiment
systematically records a lower dark current than predicted by the
control lifetimes.  This can be explained in a variety of ways by the
model.  Longer lifetimes in the wells linearly decrease the dark
current until the well lifetime is much longer than in the barrier. 
However there is no reason why the lifetimes in the QW should be greater
than in bulk material.

The approach we have investigated is the possibility of a lower QFL
separation $\delta E_{f}$ in the quantum wells. We find that a narrowing
of the QWFL separation in the wells by a factor symmetrical for holes and
electrons models all samples well, as shown by the dashed lines in graphs
\ref{iv2027} to \ref{iv951} despite the fact that the barrier bandgap varies
significantly, and that one sample has
50 wells.

This reaffirms work \cite{nelsonjap97} on the QFL separation
which was found to be significantly smaller in SQW samples at 10meV


\section{Conclusions}

The QWSC benefits from an increase in photogeneration in the wells
which is counterpointed by recombination in the lower bandgap well
regions.  In order to study which effect is greater we study
photocurrent and dark current and express the modifications to dark 
and photocurrent in terms of the quantum well density of states. The 
photocurrent from the wells is determined with no free parameters to 
good accuracy.

The dark current for the control homojunctions is well understood. 
Modelling these gives us an estimate of carrier lifetimes in the SCR,
together with verifying the transport parameters used in the $QE$
calculation at the onset of ideal diode behaviour, if present.

The QWSC dark current depends on four lifetimes. We reduce 
this to two by assuming that hole and electron non radiative lifetimes
are equal. In the absence of direct measurements we derive the barrier 
lifetimes from controls with the barrier composition. Similarly, 
controls with the well composition set an upper limit on the well 
lifetimes employed.

We see a systematic overestimation of the dark current, indicating
that the well lifetimes in a QWSC system are longer than expected by
one order of magnitude to a first approximation.  This can be
explained in terms of longer well lifetimes or a smaller QFL
separation in the wells.  This strongly suggests that these structures
have the potential to be efficient photoconverters.

Finally we note that this formulation of the MQW dark current problem shows
shortcomings because of the approximations made. This is particularly
noticeable in the case of single well samples. A full treatment needs
a self consistent approach to the Poisson equation.

This would be noticeable for single quantum well samples where the position
of the quantum wells and the background doping is important.  We note
however that the treatment applies reliably to MQW systems studied
here since dark current is determined mainly by the balance between QW
and barrier material, in that on average the position of the wells is
not critical. This is borne out by the range of \gaasalgaas\ samples
examined.

\end{document}